\documentclass[preprint,prd,superscriptaddress,showpacs]{revtex4}
\usepackage{graphicx}
\usepackage{dcolumn}
\usepackage{bm}
\newcommand{\p}{\partial}
\newcommand{\pslash}{p\kern-1ex /}
\newcommand{\lslash}{l\kern-1ex /}
\newcommand{\kslash}{k\kern-1ex /}
\newcommand{\dslash}{\p\kern-1.2ex /}
\newcommand{\Dslash}{{\cal D}\kern-1.5ex /}
\newcommand{\Aslash}{A\kern-1.2ex /}

\newcommand{\tr}{{\rm tr}}
\newcommand{\Tr}{{\rm Tr}}

\newcommand{\bea}{\begin{eqnarray}}
\newcommand{\eea}{\end{eqnarray}}
\newcommand{\vol}{\Omega}

\newcommand{\BAN}{\begin{eqnarray*}}
\newcommand{\EAN}{\end{eqnarray*}}
\begin{document}

\newcommand{\NTU}{
  Physics Department, 
  Center for Quantum Science and Engineering,  
  and Center for Theoretical Sciences, 
  National Taiwan University, Taipei~10617, Taiwan  
}

\newcommand{\RCAS}{
  Research Center for Applied Sciences, Academia Sinica,
  Taipei~115, Taiwan
}

\preprint{NTUTH-08-505D}
 
\title{Topological susceptibility in 2+1 flavors lattice QCD 
       with domain-wall fermions}

\author{Ting-Wai~Chiu}
\affiliation{\NTU}

\author{Tung-Han~Hsieh}
\affiliation{\RCAS}

\author{Po-Kai~Tseng}
\affiliation{\NTU}

\collaboration{for the TWQCD Collaboration}
\noaffiliation


\begin{abstract}

  We measure the topological charge and its fluctuation 
  for the gauge configurations generated by the RBC and UKQCD Collaborations  
  using 2+1 flavors of domain-wall fermions on the $ 16^3 \times 32 $ 
  lattice ($ L \simeq 2 $ fm) with length 16 in the fifth dimension 
  at the inverse lattice spacing $ a^{-1} \simeq 1.62 $ GeV. 
  From the spectral flow of the Hermitian 
  operator $ H_w ( 2 + \gamma_5 H_w)^{-1} $, we obtain the topological 
  charge $ Q_t $ of each gauge configuration in the three ensembles with 
  light sea quark masses $ m_q a = 0.01 $, $ 0.02 $, and $ 0.03 $, 
  and with the strange quark mass fixed at $ m_s a = 0.04 $.
  From our result of $ Q_t $, we compute the topological susceptibilty 
  $ \chi_t = \langle Q_t^2 \rangle /\vol $,  
  where $ \vol $ is the volume of the lattice.
  In the small $m_q$ regime, our result of $\chi_t $ agrees with 
  the chiral effective theory. Using the formula 
  $ \chi_t = \Sigma / (m_u^{-1} + m_d^{-1} + m_s^{-1}) $
  by Leutwyler-Smilga, we obtain the chiral condensate 
  $\Sigma^{\overline{\mathrm{MS}}}(\mathrm{2~GeV})
  =[\mathrm{259(6)(9)~MeV}]^3 $. 

\end{abstract}

\maketitle

\section{Introduction}

In Quantum Chromodynamics (QCD), the topological susceptibility
($ \chi_t $) is the most crucial quantity to measure the
topological charge fluctuation of the QCD vacuum,
which plays an important role in breaking the $ U_A(1) $ symmetry.
Theoretically, $ \chi_t $ is defined as
\bea
\label{eq:chi_t}
\chi_{t} = \int d^4 x  \left< \rho(x) \rho(0) \right>, 
\eea
where 
\bea
\label{eq:rho}
\rho(x) = \frac{1}{32 \pi^2} \epsilon_{\mu\nu\lambda\sigma}
                             \tr[ F_{\mu\nu}(x) F_{\lambda\sigma}(x) ], 
\eea
is the topological charge density 
expressed in term of the matrix-valued field tensor $ F_{\mu\nu} $.
With mild assumptions, Witten \cite{Witten:1979vv} and
Veneziano \cite{Veneziano:1979ec}
obtained a relationship between the topological susceptibility
in the quenched approximation and the mass of $ \eta' $ 
meson (flavor singlet) in unquenched QCD with $ N_f $ degenerate flavors, 
namely, 
\BAN
\chi_t(\mbox{quenched}) = \frac{f_\pi^2 m_{\eta'}^2}{4 N_f}, 
\EAN
where $ f_\pi = 131 $ MeV, the decay constant of pion.
This implies that the mass of $ \eta' $ is essentially due to
the axial anomaly relating to non-trivial topological charge
fluctuations, which can turn out to be nonzero even in the chiral limit,
unlike those of the (non-singlet) approximate Goldstone bosons.

Using the Chiral Perturbation Theory (ChPT), 
Leutwyler and Smilga \cite{Leutwyler:1992yt}
obtained the following relation in the chiral limit  
\bea
\label{eq:LS}
\chi_t = \frac{\Sigma}
         {\left( \frac{1}{m_u} +\frac{1}{m_d} +\frac{1}{m_s} \right)}
+ {\cal O}(m_u^2), \hspace{4mm} (N_f = 2+1), 
\eea
where $ m_u$, $m_d$, and $m_s$ are the quark masses, 
and $ \Sigma $ is the chiral condensate.
This implies that in the chiral limit ($ m_u \to 0 $) 
the topological susceptibility is suppressed due to internal quark loops.
Most importantly, (\ref{eq:LS}) provides a viable way
to extract $ \Sigma $ from $ \chi_t $ in the chiral limit.

From (\ref{eq:chi_t}), one obtains
\bea
\label{eq:chit_Qt}
\chi_t = \frac{\left< Q_t^2 \right>}{\Omega}, \hspace{4mm}
Q_t \equiv  \int d^4 x \rho(x), 
\eea
where $ \Omega $ is the volume of the system, and
$ Q_t $ is the topological charge (which is an integer for QCD).
Thus, one can determine $ \chi_t $ by counting the number of
gauge configurations for each topological sector.  
Furthermore, we can also obtain the second normalized cumulant 
\bea 
\label{eq:c4}
c_4 = -\frac{1}{\vol} \left[   \langle Q_t^4 \rangle
                            -3 \langle Q_t^2 \rangle^2 \right], 
\eea
which is related to the leading anomalous contribution to 
the $ \eta'-\eta' $ scattering amplitude in QCD, as well as the 
dependence of the vacuum energy on the vacuum angle $ \theta $.     
(For a recent review, see for example, Ref. \cite{Vicari:2008jw} 
and references therein.)

However, for lattice QCD, it is difficult to extract $ \rho(x) $ 
and $ Q_t $ unambiguously from the gauge link variables, due to 
their rather strong fluctuations.

To circumvent this difficulty, one may consider
the Atiyah-Singer index theorem 
\cite{Atiyah:1968mp}
\bea
\label{eq:AS_thm}
Q_t = n_+ - n_- = \mbox{index}({\cal D}), 
\eea
where $ n_\pm $ is the number of zero modes of the massless Dirac
operator $ {\cal D} \equiv \gamma_\mu ( \partial_\mu + i g A_\mu) $
with $ \pm $ chirality. 

For lattice QCD with exact chiral symmetry, it is well-known that 
the overlap Dirac operator \cite{Neuberger:1997fp,Narayanan:1995gw} 
in a topologically non-trivial gauge background
possesses exact zero modes (with definite chirality) satisfying
the Atiyah-Singer index theorem. Thus we can obtain the topological 
charge from the index of the overlap Dirac operator.  
Writing the overlap Dirac operator as
\bea
\label{eq:overlap}
D = m_0 \left( 1 + \gamma_5 \frac{H_w}{\sqrt{H_w^2}} \right),  
\eea
where $ H_w $ is the standard Hermitian Wilson operator with negative
mass $ -m_0 $ ($ 0 < m_0 < 2 $), then its index is 
\bea
\label{eq:index_overlap}
 \mbox{index}(D)  
= \Tr \left[ \gamma_5 \left( 1 - \frac{D}{2m_0} \right) \right] 
= -\frac{1}{2} \Tr \left( \frac{H_w}{\sqrt{H_w^2}} \right) = n_+ - n_- 
= Q_t,  
\eea
where $ \Tr $ denotes trace over Dirac, color and lattice spaces.

Obviously, from (\ref{eq:index_overlap}), we have 
\bea
\label{eq:ind_D}
\mbox{index}( D ) = n_+ - n_- 
= -\frac{1}{2} \Tr \left( \frac{H_w}{\sqrt{H_w^2}} \right) 
= \frac{1}{2} (h_{-} - h_{+}), 
\eea
where $ h_{+} (h_{-}) $ is the number of positive (negative)
eigenvalues of the hermitian Wilson-Dirac operator $ H_w $.
However, one does not need to obtain all eigenvalues of $ H_w $
in order to know how many of them are positive or negative.
The idea is simple. Since $ H_w $ has equal number of positive and
negative eigenvalues for $ m_0 \le 0 $, then one can just focus on those
low-lying (near zero) eigenmodes of $ H_w $, and see whether any of
them crosses zero from positive to negative, or vice versa, when $ m_0 $
is scanned from zero up to the value used in the definition of $ D $. 
From the net number of crossings, one can obtain the index of $ D $. 
This is the spectral flow method used in Ref. \cite{Narayanan:1995gw}
to obtain the index of the overlap Dirac operator. 
We also applied the spectral flow method to obtain the index 
of the overlap Dirac operator and to determine the topological
susceptibility in quenched QCD \cite{Chiu:2003iw}.
In this paper, we extend our previous studies to unquenched QCD. 

\begin{figure}[htb]
\begin{center}
\includegraphics[height=10cm,width=14cm]{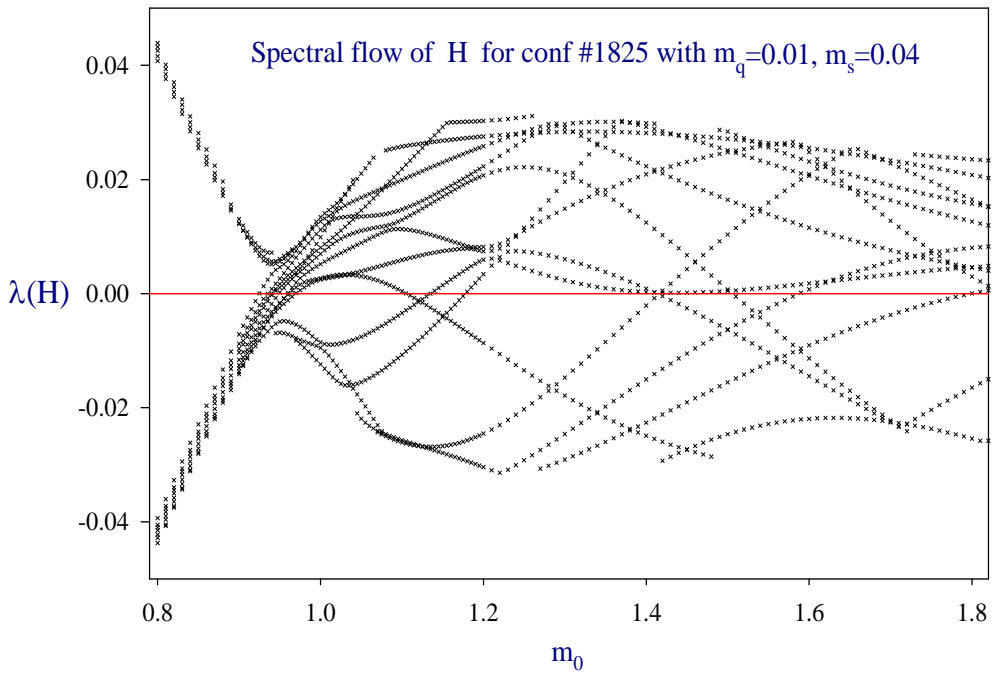}
\caption{The spectral flow of 12 lowest-lying eigenvalues of $ H $ for the
gauge configuration number 1825 in the ensemble ($m_q = 0.01 $, $m_s = 0.04 $). 
There are 10 net crossings from negative to positive, 
so the index is $-10$.}
\label{fig:specflow_m001_1825}
\end{center}
\end{figure}

For the conventional domain-wall fermion \cite{Kaplan:1992bt,Shamir:1993zy}, 
its effective 4-dimensional Dirac operator 
is 
\bea
\label{eq:D_dwf}
D = \frac{m_0 (2-m_0)}{2} \left( 1 + \gamma_5 \frac{H}{\sqrt{H^2}} \right),  
\eea
where 
\bea
H = H_w ( 2 + \gamma_5 H_w )^{-1}. 
\eea
Thus, we can obtain the topological charge of a gauge field configuration 
from the spectral-flow of $ H $ as a function of $ m_0 $.
Obviously, it is much more computationally intensive to project 
the low-lying eigenvalues of $H$ than those of $ H_w $, due to the  
extra inverse operator $ (2+\gamma_5 H_w)^{-1} $.

In this paper, we use the spectral flow of $ H $ to determine
the topological charge of the gauge configurations
(http://lattices.qcdoc.bnl.gov/) generated by the 
RBC and UKQCD Collaborations using 2+1 flavors of domain-wall
fermions on the $ 16^3 \times 32 $ lattice ($L \simeq 2 $ fm) with 
length 16 in the fifth dimension at the inverse lattice spacing
$ a^{-1} \simeq 1.62(4) $ GeV \cite{Allton:2007hx}.
There are three ensembles of gauge configurations with
light sea quark masses $ m_q a = 0.01 $, $ 0.02 $ and $ 0.03 $,
and with the strange quark mass fixed at $ m_s a = 0.04 $.
For the ensmeble with $ m_q a = 0.01 $, we pick one configuration every
5 configurations, from configurations numbering from 0020 to 4015.
Thus we have 800 configurations with $ m_q a = 0.01 $.
Similarly, for $ m_q a = 0.02 $,
we pick 809 configurations from configurations numbering from 0005 to 4045,
and for $ m_q a = 0.03 $,
we pick 717 configurations from configurations numbering from 4020 to 7600.

In Fig. \ref{fig:specflow_m001_1825}, we plot the spectral flow 
of 12 lowest-lying (near zero) eigenvalues of $ H(m_0) $ in the interval
$ 0.8 \le m_0 \le 1.8 $, for the gauge configuration number 1825 in 
the ensemble with $ m_q = 0.01 $ and $ m_s = 0.04 $.
In this case, the net crossings from negative to positive is 10,
so the index is $-10$.
In general, it may happen that there are some intriguing eigenvalues 
lying very close to zero 
(e.g., the one around $ m_0 = 1.45 $ in Fig. \ref{fig:specflow_m001_1825}). 
Thus, with a coarse scan in $ m_0 $, it may not be able to determine 
whether they actually cross zero or not. 
These ambiguities can only be resolved by
tracing them closely at a finer resolution in $ m_0 $.
Obviously, it is a very tedious job to determine the topological 
charges of 2,326 gauge configurations
via the spectral flow of $ H = H_w ( 2 + \gamma_5 H_w )^{-1} $.

\begin{figure}[htb]
\begin{center}
\begin{tabular}{@{}ccc@{}}
\includegraphics*[height=8cm,width=5.5cm]{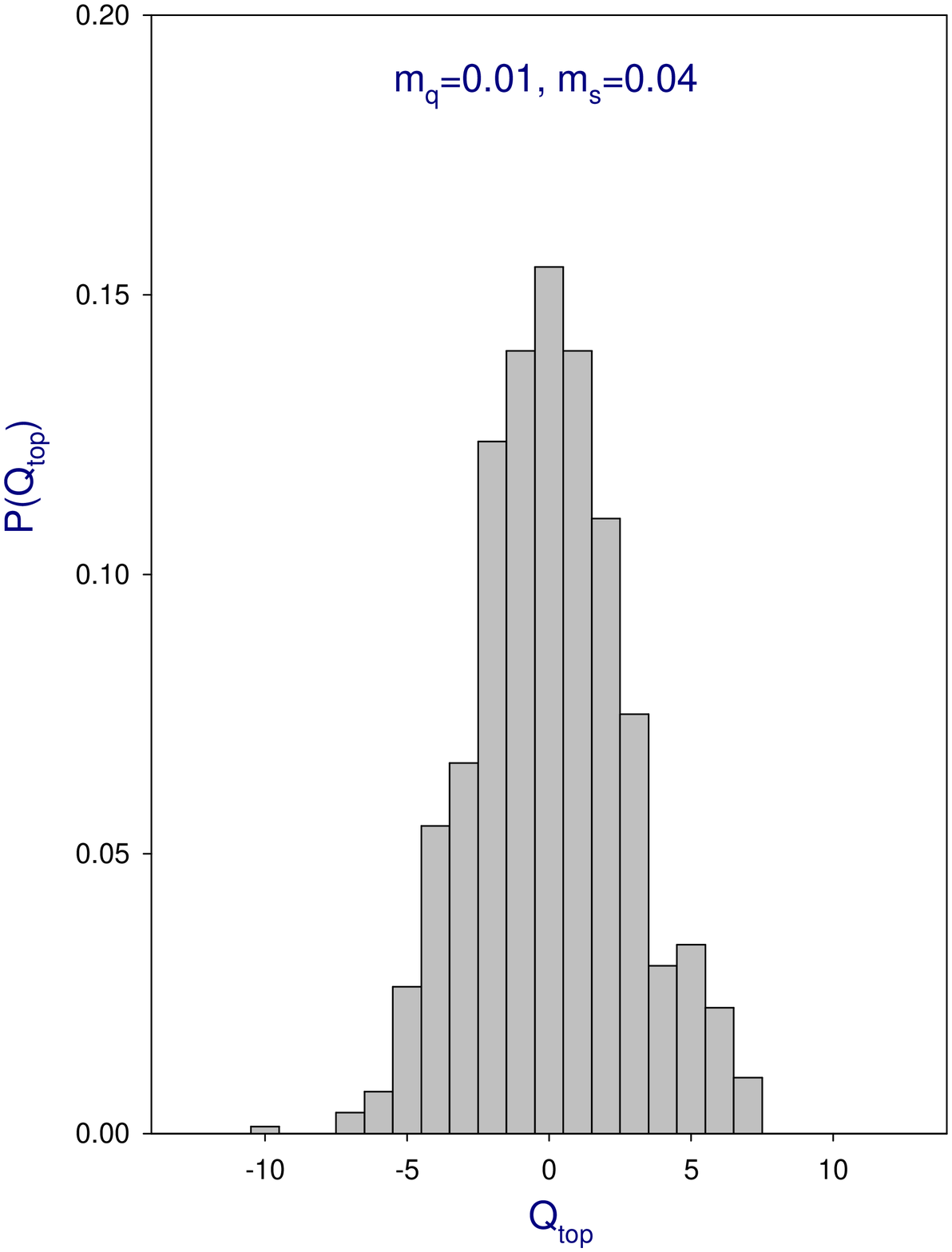}
&
\includegraphics*[height=8cm,width=5.5cm]{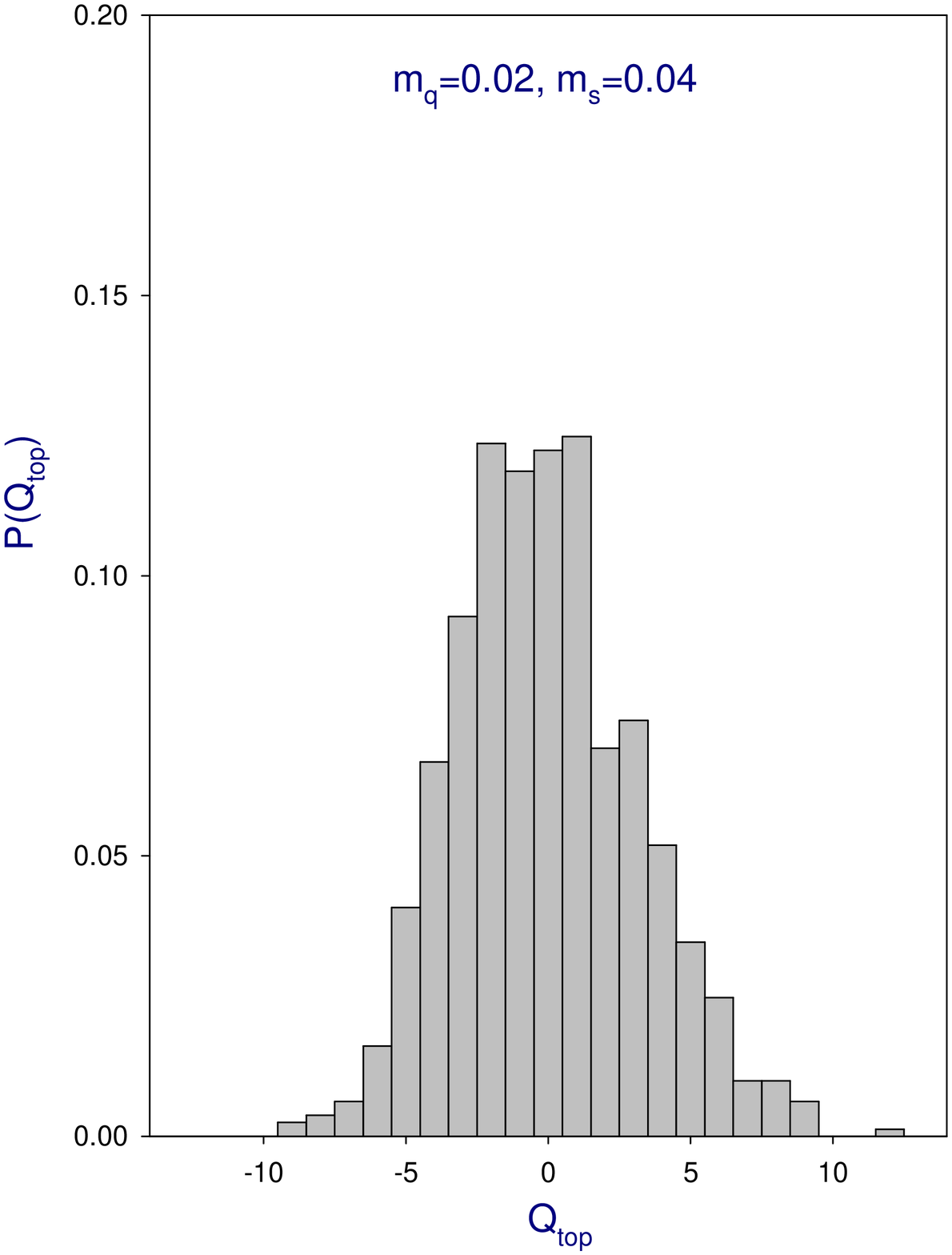}
&
\includegraphics*[height=8cm,width=5.5cm]{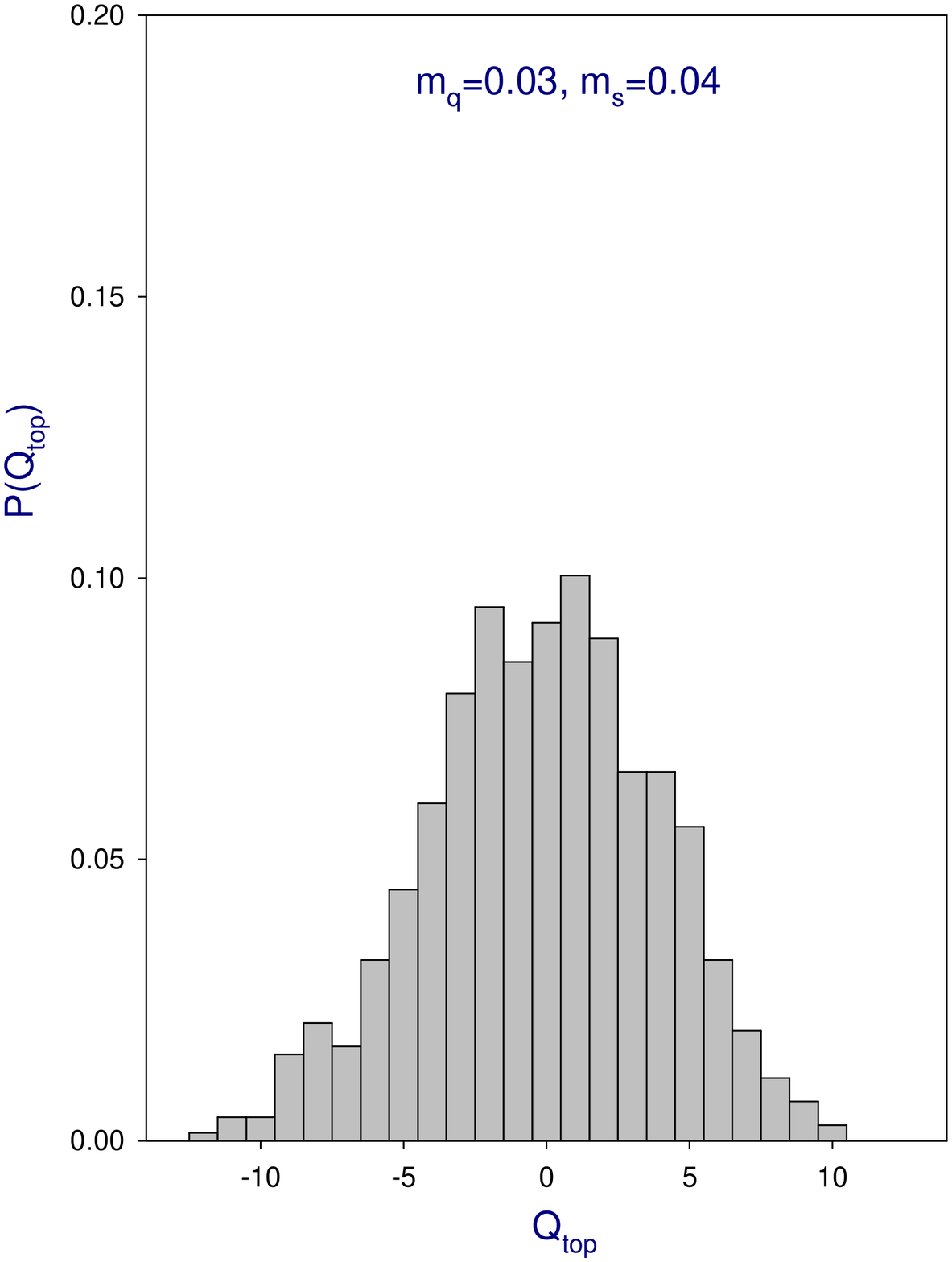}
\end{tabular}
\caption{Histogram of topological charge distribution 
for $ m_q a =0.01, 0.02 $, and $ 0.03 $ respectively.    
}
\label{fig:Q_hist}
\end{center}
\end{figure}

\section{Results}

In Fig.~\ref{fig:Q_hist}, we plot the histogram of topological charge    
distribution for $ m_q a =0.01, 0.02 $, and $ 0.03 $ respectively. 
Evidently, the probability distribution of $ Q_t $ becomes more 
sharply peaked around $ Q_t = 0 $ as the light sea quark mass $ m_q $ 
gets smaller. Our results of topological susceptibility $\chi_t $, 
and the second normalized cumulant $ c_4 $, together with their
ratios $ c_4 / \chi_t $, and $ c_4 / ( 2 \chi_t^2 \vol) $ 
are listed in Table \ref{tab:chit_c4_mq}.
The error is estimated using the jackknife method with bin size of
20 configurations for $ m_q = 0.01 $ and $ 0.02 $, and 13 
configurations for $ m_q = 0.03 $, 
with which the statistical error saturates.

Evidently, the statistical error of the topological susceptibility 
is about 10\%, while that of $ c_4 $ is very large due to low statistics. 
Therefore, one cannot draw any conclusions from our result of $ c_4 $,  
as well as from the ratio $ c_4 / \chi_t $. Interestingly, our result of 
$ c_4 /(2 \chi_t^2 \vol) $ is consistent with that  
in Ref. \cite{Chiu:2008kt}, which is obtained 
from the plateaus (at large time separation)
of the 2-point and 4-point time-correlation functions of the
flavor-singlet pseudoscalar meson $\eta'$ in a fixed global 
topology with $ Q_t = 0 $.

\begin{table}[th]
\begin{center}
\begin{tabular}{|c|cccc|}
\hline
$ m_q a $ & $ \chi_t $ & $ c_4 $ & $ c_4/\chi_t $ & $ c_4/(2\chi_t^2 \vol) $ \\
\hline
0.01    & $ 5.51(62) \times 10^{-5} $ & $ -1.88(8.33) \times 10^{-5} $
        & $ -0.34(1.55) $ & $ -0.0239(1110) $       \\
0.02    & $ 7.74(75) \times 10^{-5} $ & $ -4.99(16.18) \times 10^{-5} $
        & $ -0.64(2.15) $ & $ -0.0317(1088) $      \\
0.03    & $ 1.23(11) \times 10^{-4} $ & $ 4.69(3.46) \times 10^{-4} $
        & $ 3.80(2.57) $ & $ 0.1172(834) $ \\
\hline
\end{tabular}
\end{center}
\caption{
The topological susceptibility $ \chi_t $,
the second normalized cumulant $ c_4 $, and their ratios $ c_4/\chi_t $,
and $ c_4 /(2 \chi_t^2 \vol) $, versus the sea quark masses, for
$ N_f = 2+1 $ lattice QCD with domain-wall fermions.
}
\label{tab:chit_c4_mq}
\end{table}

\begin{figure}[tb]
\centering
\includegraphics[width=12cm,height=8cm]
                {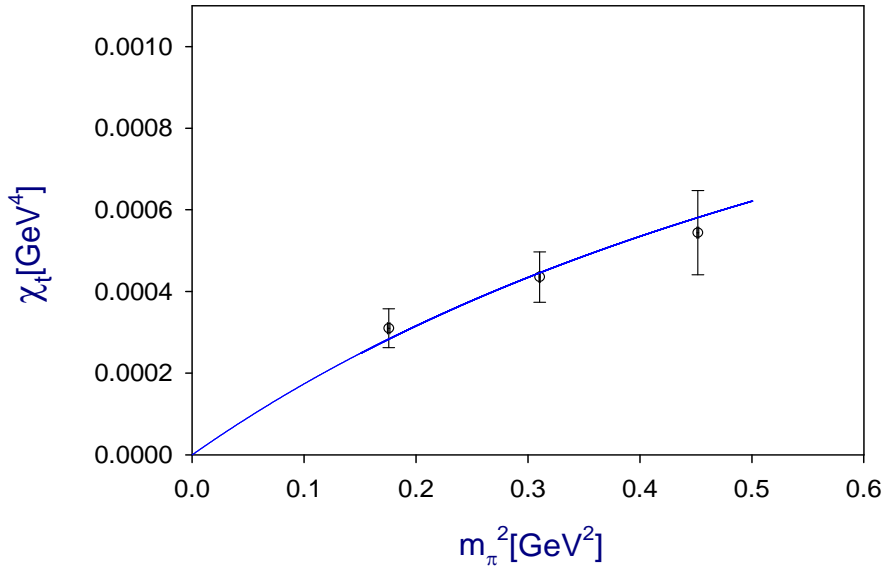}
\caption{The topological susceptibility $ \chi_t $
         versus $ m_\pi^2 $ for 2+1 flavors lattice QCD with 
         domain-wall fermions.}
\label{fig:chit_mpi2_nf2p1}
\end{figure}

In Fig \ref{fig:chit_mpi2_nf2p1}, 
we plot our data of $ \chi_t $ versus $ m_\pi^2 $, 
where the pion mass $ m_\pi a $ and the inverse lattice spacing $ a^{-1} $ 
are determined by the RBC and UKQCD Collaborations \cite{Allton:2007hx}. 

For three flavors with $ m_u = m_d $, we may use the partial
conservation of the axial current (PCAC) relation $ m_{PS}^2 = C m_q $
to transcribe the Leutwyler-Smilga relation (\ref{eq:LS}) to 
\bea
\label{eq:chit_LS_PCAC}
\chi_t = \frac{\Sigma' m_{\pi}^2}
              {2 \left[1 + \frac{m_\pi^2}{2 m_{PS}^2(m_s)} \right] },  
\eea
where $ m_{PS}(m_s) $ is the mass of the pseudoscalar meson 
with strange quarks, and $ \Sigma' = \Sigma C^{-1} $.  
The data points of $ \chi_t $ are well fitted by  
the ChPT formula (\ref{eq:chit_LS_PCAC}) 
with $ \Sigma' = 0.0039(2) [\mathrm{GeV}^2] $.
The fitted curve is plotted as the solid line in      
Fig \ref{fig:chit_mpi2_nf2p1}.  
Using $ C = 7.4034(1986) [\mathrm{GeV}] $ determined by the
RBC and UKQCD Collaborations \cite{Allton:2007hx}, 
we obtain $ \Sigma = 0.0288(16) [\mathrm{GeV}^3] $.

In order to convert $\Sigma$ to that in the
$\overline{\mathrm{MS}}$ scheme, we use the renormalization factor 
$Z_s^{\overline{\mathrm{MS}}}(\mathrm{2~GeV}) = 0.604(18)(55) $ 
which is determined by the RBC and UKQCD Collaborations \cite{Aoki:2007xm}, 
employing the non-perturbative renormalization technique
through the RI/MOM scheme \cite{Martinelli:1994ty}.
Then the value of $ \Sigma $ is transcribed to
$$
\Sigma^{\overline{{\mathrm{MS}}}}(\mathrm{2~GeV})
  =[\mathrm{259(6)(9)~MeV}]^3,  
$$ 
which is in good agreement with the results extracted from $ \chi_t $
in 2+1 flavors QCD \cite{Chiu:2008kt} and 2 flavors QCD \cite{Aoki:2007pw}, 
as well as that obtained from the low-lying eigenvalues 
in the $\epsilon$-regime \cite{Fukaya:2007yv}.
The errors represent a combined statistical error
($a^{-1}$ and $Z_s^{\overline{\mathrm{MS}}}$) and
the systematic error respectively.
Since the calculation is done at a single lattice spacing,
the discretization error cannot be quantified reliably, but
we do not expect much larger error because the domain-wall fermion 
action is free from $O(a)$ discretization effects.

\section{Concluding remark}
  
In this paper, we have obtained the topological susceptibility $\chi_t$
and the second normalized cumulant $ c_4 $, in 2+1 flavors lattice QCD 
with domain-wall fermions.
The expected sea quark mass (pion mass square) dependence of $\chi_t$ 
from ChPT is clearly observed. However, our statistics of $ \sim 800 $ 
configurations are insufficient to determine $ c_4 $ 
with a reasonably small error. 

Finally, it is interesting to compare the topological susceptibility
obtained from the global topological charge (this work) with that 
extracting from the plateaus (at large time separation)
of the 2-point and 4-point time-correlation functions of the
flavor-singlet pseudoscalar meson $\eta'$ with fixed topology 
\cite{Chiu:2008kt}. 
We find that both methods give $ \chi_t $ in agreement with the 
chiral effective theory. 
Furthermore, the chiral condensates extracted from 
both sets of $ \chi_t $ are also in good agreement.  
Next, we turn to the second normalized cumulant $ c_4 $. 
If we try to measure it with the global topological charge, it 
would require more than 10,000 configurations in order to 
pin down the statistical error less than 10\%. 
On the other hand, if we measure $ c_4 $ with the correlation function 
of topological charges in sub-volumes (with fixed global topology), 
we would expect that it only requires about 1,000 configurations 
in order to achieve a level of 10\% statistical error. 
To conclude, it is interesting to see that there are more than 
one viable options to obtain the topological susceptibility and the higher 
normalized cumulants, in lattice QCD with exact chiral symmetry.

\begin{acknowledgments}
  We thank the RBC and UKQCD Collaborations for releasing their 
  gauge configurations to the public at the website 
  http://lattices.qcdoc.bnl.gov/.
  Our numerical computations are mostly performed  
  at the National Center for High Performance Computing, 
  and the Computing Center at National Taiwan University. 
  This work is supported in part by the National Science Council 
   (No.~NSC96-2112-M-002-020-MY3, 
        NSC96-2112-M-001-017-MY3, and NSC97-2119-M-002-001),   
    and also NTU-CQSE (No.~97R0066-69). 
\end{acknowledgments}

\end{document}